\documentclass{pasj00}

\begin{document}
\SetRunningHead{M. Suzuki et al.}{Discovery of a New X-ray Burst/Millisecond Accreting Pulsar HETE~J1900.1-2455}
\Received{2006/00/00}
\Accepted{0000/00/00}

\title{Discovery of a New X-Ray Burst/Millisecond Accreting
	Pulsar HETE~J1900.1-2455}

\author{%
  Motoko \textsc{Suzuki}\altaffilmark{1}
  Nobuyuki \textsc{Kawai}\altaffilmark{2,1}
  Toru \textsc{Tamagawa}\altaffilmark{1}
  Atsumasa \textsc{Yoshida}\altaffilmark{3,1}\\
  Yujin E. \textsc{Nakagawa}\altaffilmark{3}
  Kaoru \textsc{Tanaka}\altaffilmark{3}
  Yuji \textsc{Shirasaki}\altaffilmark{4}
  Masaru \textsc{Matsuoka}\altaffilmark{5}\\
  George R. \textsc{Ricker}\altaffilmark{6}
  Roland \textsc{Vanderspek}\altaffilmark{6}
  Nat \textsc{Butler}\altaffilmark{6,7}
  Donald Q. \textsc{Lamb}\altaffilmark{8}
  Carlo \textsc{Graziani}\altaffilmark{8}\\
  Graziella \textsc{Pizzichini}\altaffilmark{9}
  Rie \textsc{Sato}\altaffilmark{2}
  Makoto \textsc{Arimoto}\altaffilmark{2}
  Jun'ichi \textsc{Kotoku}\altaffilmark{2}\\
  Miki \textsc{Maetou}\altaffilmark{3}
  and
  Makoto \textsc{Yamauchi}\altaffilmark{10}
}
\altaffiltext{1}{The Institute of Physical and Chemical Research, 2-1 Hirosawa, Wako, Saitama 351-0198}
\email{motoko@crab.riken.jp}
\altaffiltext{2}{Department of Physics, Tokyo Institute of Technology, \\2-12-1 Ookayama, Meguro-ku, Tokyo, 152-8551}
\altaffiltext{3}{Department of Physics and Mathematics, Aoyama Gakuin University, \\5-10-1 Fuchinobe, Sagamihara, Kanagawa 229-8558}
\altaffiltext{4}{National Astronomical Observatory, 2-21-1 Osawa, Mitaka, Tokyo, 181-8588}
\altaffiltext{5}{Tsukuba Space Center, JAXA, 2-1-1 Sengen, Tsukuba, Ibaraki, 305-8505}
\altaffiltext{6}{Center for Space Research, MIT, 70 Vassar Street, Cambridge, Massachusetts, 02139, USA}
\altaffiltext{7}{Space Sciences Laboratory, University of California at Berkeley, \\Berkeley, California, 94720-7450, USA}
\altaffiltext{8}{Department of Astronomy and Astrophysics, University of Chicago, \\5640 South Ellis Avenue, Chicago, Illinois 60637, USA}
\altaffiltext{9}{INAF/IASF Bologna, Via Gobetti 101, 40129 Bologna, Italy}
\altaffiltext{10}{Faculty of Engineering, Miyazaki University, Gakuen Kibanadai Nishi, Miyazaki, 889-2192}


\maketitle

\begin{abstract}
A class of low-mass X-ray binary sources are known to be both
X-ray burst sources and millisecond pulsars at the same time.
A new source of this class was discovered by High Energy Transient
Explorer 2 (HETE-2) on 14 June 2005 as a source of type-I X-ray
bursts, which was named HETE J1900.1-2455. 
Five  X-ray bursts from HETE J1900.1-2455 were observed during the
summer of 2005.  
The time resolved spectral analysis of these bursts have revealed that
their spectra are consistent with the blackbody radiation throughout
the bursts.  The bursts show the indication of radius expansion.
The bolometric flux remains almost constant during the photospheric
radius expansion while blackbody temperature
dropped during the same period.  Assuming that the flux reached to the
Eddington limit on a standard 1.4 solar mass neutron star with a helium
atmosphere, we estimate the distance to the source to be $\sim$ 4 kpc.
\end{abstract}

\section{Introduction}

An X-ray burst source HETE J1900.1-2455 was discovered
by High Energy Transient Explorer 2 (HETE-2) on 14 June 
2005.  The position of the source was distributed through the
Gamma ray bursts Coordinates Network circular
\citep{2005GCN..3548....1A} and The Astronomer's Telegram
\citep{2005ATel..516....1V}.

The follow-up observations were performed
by the Rossi X-ray Timing Explorer (RXTE) on June 16 and 17 in 2005
with exposures of 1.14 ks and 5.46 ks respectively.  An X-ray pulsar
with a spin frequency of 377.3 Hz was found at
R.A. = \timeform{19h 00m 13s}, Dec. = \timeform{-24D 54' 44"} (J2000)
and this position was consistent with 
the HETE error circle.
The flux of the pulsar was 6.6 $\times$ 10$^{-10}$ erg 
cm$^{-2}$ s$^{-1}$ in 2$-$20 keV
\citep{2005ATel..523....1M,2005ATel..525....1M,2006ApJ...638..963K}.
\citet{2005ATel..526....1F} observed the field of the SXC error
circle with the Robotic Palomar 60-inch Telescope with
R-band and i-band from 07:27 to 07:57 UT on 18 June 2005,
and detected a previously unknown object at
R.A. =\timeform{19h00m08s.65}, Dec =\timeform{-24D55'13".7} (J2000).
The brightness of the counterpart is R $\sim$ 18.4 mag.

\begin{figure*}
 \begin{center}
   \FigureFile(175mm,){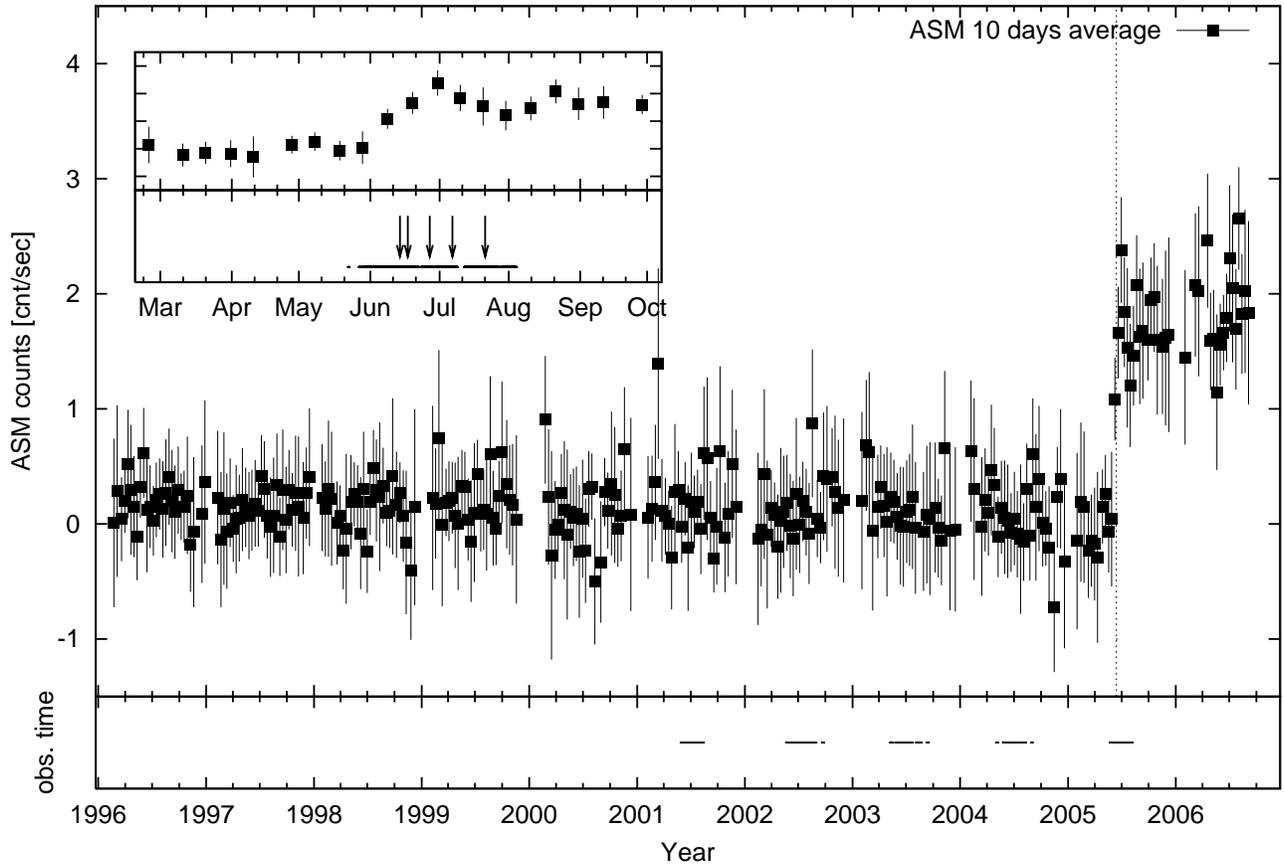}
 \end{center}
 \caption{
          The light curve of HETE J1900.1-2455 observed by RXTE/ASM
           (upper panel),
          and the observation time coverage of HETE-2/WXM, where the
          time intervals when HETE J1900.1-2455 was in the 
          HETE-2/WXM field of view are shown with dots (lower panel).
           The time of the first burst detected by HETE-2
          is shown with the vertical dotted line, which coincides with
          the time of the sudden brightening.
          The ASM light curve and the WXM coverage in 2005 are
          enlarged in the insets with the time of the bursts shown
          with arrows.
 }
 \label{fig.asm}
\end{figure*}

HETE J1900.1-2455 is the seventh known accretion-powered millisecond
pulsar.  The sources of this class have some common properties.
Their orbital periods are distributed between 40 min and 120 min
\citep{2006ApJ...638..963K}, spectra above 15 keV were described
by power-law models with exponential cut-off 
(for example \cite{2005AA...436..647F,2005AA...444...15F}), 
while low energy spectra need blackbody and/or disk blackbody
components
(for example \cite{2005MNRAS.359.1261G,2003ApJ...587..754J}).
In fact, through the observation of HETE J1900.1-2455 by RXTE,
power-law and blackbody spectral features are found
by \citet{2006ApJ...638..963K} and a hint of high energy emission
is reported by \citet{2005ATel..590....1G}.
The upper panel of figure \ref{fig.asm} shows the light curve of
HETE J1900.1-2455 observed by the All-Sky Monitor (ASM) on RXTE
\footnote{http://xte.mit.edu/ASM\_lc.html}%
. The vertical dashed line indicates the time of the first burst.
The persistent flux of HETE J1900.1-2455 suddenly
brightened at the time of the first burst and
remains almost constant after that.
Since the orbital period of the source is 83.3 min
\citep{2006ApJ...638..963K}, the sudden brightening is not likely
to be related to the orbital motion of the source.
In contrast, the other accretion-powered millisecond pulsars
are transients with typical decay time of a few weeks.
According to \citet{2006ApJ...638..963K}, the mass of companion is
probably between 0.016 $M_{\Sol}$ and 0.07 $M_{\Sol}$.
\citet{2006MNRAS.366L..31K} suggested the possibility that the
companion was born as an object of a small mass, such as a brown dwarf
or a planet, though the detailed path of evolution of the system is not
clear.

HETE-2 detected five X-ray
bursts from HETE J1900.1-2455 in total. The time of the bursts are
shown with arrows in the insets of figure \ref{fig.asm}.
In this paper we report the timing and spectral properties of
these five X-ray bursts from HETE J1900.1-2455 observed during
the summer of 2005.

\section{Observation and data analysis}

\subsection{Localization}
The French Gamma Telescope (FREGATE), the Wide-field X-ray Monitor
(WXM), and the Soft X-ray Camera (SXC) instruments on board
HETE-2 detected an X-ray burst (trigger ID H3804) at 11:22 UT on
14 June 2005.  

\begin{figure}
 \begin{center}
    \FigureFile(80mm,){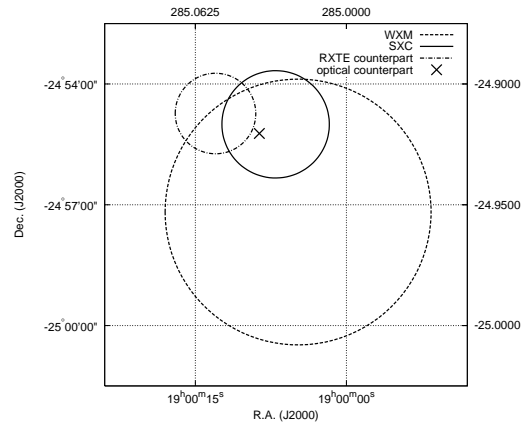}
 \end{center}
 \caption{The Skymap of HETE J1900.1-2455 localization.  The circles
 with dashed and solid lines are the error circles derived
 with WXM and SXC data .   The dash-dotted circle is the location 
 of the position of the X-ray pulsar \citep{2005ATel..525....1M}.
 The position of the optical counterpart \citep{2005ATel..526....1F}
 is plotted with a cross mark.}
 \label{fig.skymap}
\end{figure}

This event was localized by the ground analysis of WXM data,
and refined by SXC data to a circle of with  \timeform{80"}-radius
centered at 
R.A. = \timeform{19h 00m 6s.4}, Dec = \timeform{-24D 54' 54".7} (J2000)
\citep{2005GCN..3548....1A}.
The Galactic coordinates of the burst is 
l=\timeform{00h45m}, b=\timeform{-12D.86}.
There was no known X-ray source at this position.
The position of H3804 did not match any of the entry in 
in the globular cluster catalog
\citep{1996AJ....112.1487H} either, while
globular clusters sometimes host X-ray burst sources. 
The results of localizations by WXM and SXC are shown
in figure \ref{fig.skymap}.
The positions of the source determined by the RXTE PCA scans \citep{2005ATel..525....1M}
and the optical counterpart candidate \citep{2005ATel..526....1F}
are also shown in the figure.

\subsection{Data analysis}
In addition to the first burst on 14 June, HETE-2/WXM detected
four more X-ray bursts from the source on 17 June, 27 June, 7 July,
and 21 July in 2005.  The burst on 14 June and 7 July were triggered
events while the others were untriggered.  
For the second burst on 17 June the information of accurate spacecraft
attitude was not available, because the optical camera system had 
been turned off about 100 seconds before the burst.  The source
position of the burst was found to be 0.6 degree off HETE J1900.1-2455 
if the attitude information of 100 seconds before the burst was used.
We consider, however, that HETE J1900.1-2455 is still
the most likely source of this burst, as 
it is possible that the spacecraft attitude
drifted 0.6 degree in 100 seconds, and there are no known X-ray sources
within comparable distances.
The time of the all events and their
incident angles are summarized in Table \ref{tab.obs}.
We note that the burst on 21 July was also detected by RXTE
\citep{2005ATel..657....1G}.

\begin{table*}
  \caption{The observation log of the five bursts observed by
    HETE-2.  The burst date and time, HETE burst ID,
    incident angle on WXM, and availability of TAG data are
    summarized in the table.  $ \theta_{x}$ and $ \theta_{y}$ are
    the projection angles measured from the vertical direction onto
    the XZ and YZ plane of the detector coordinate system.}
  \label{tab.obs}
  \begin{center}
  \begin{tabular}{cccccc}
  \hline \hline 
  Burst No.  & 1 & 2 & 3 & 4 & 5\\
  \hline 
  date (MJD)  & 14 Jun. (53535) & 17 Jun. (53538) & 27 Jun. (53548) 
          & 7 Jul. (53558) & 21 Jul. (53572) \\
  time (UT)   & 11:21:50 & 21:49:10 & 13:54:10 & 13:09:22 & 23:00:32 \\
  burst ID  & 3804     & 11663    & 11662    & 3858     & 11640    \\
  ($\theta_x$, $\theta_y$) [deg] 
  & $(4.24, \, -8.63)$ & $(3.64, \, -6.00)$ & $(-0.82, \, 3.04)$ 
          & $(18.37, \, 3.58)$ & $(14.83, \, 11.48)$ \\
  TAG data  & yes     &  no & no & yes    & yes     \\
  \hline 
  \end{tabular}
  \end{center}
\end{table*}

\begin{figure*}
 \begin{center}
    \FigureFile(140mm,){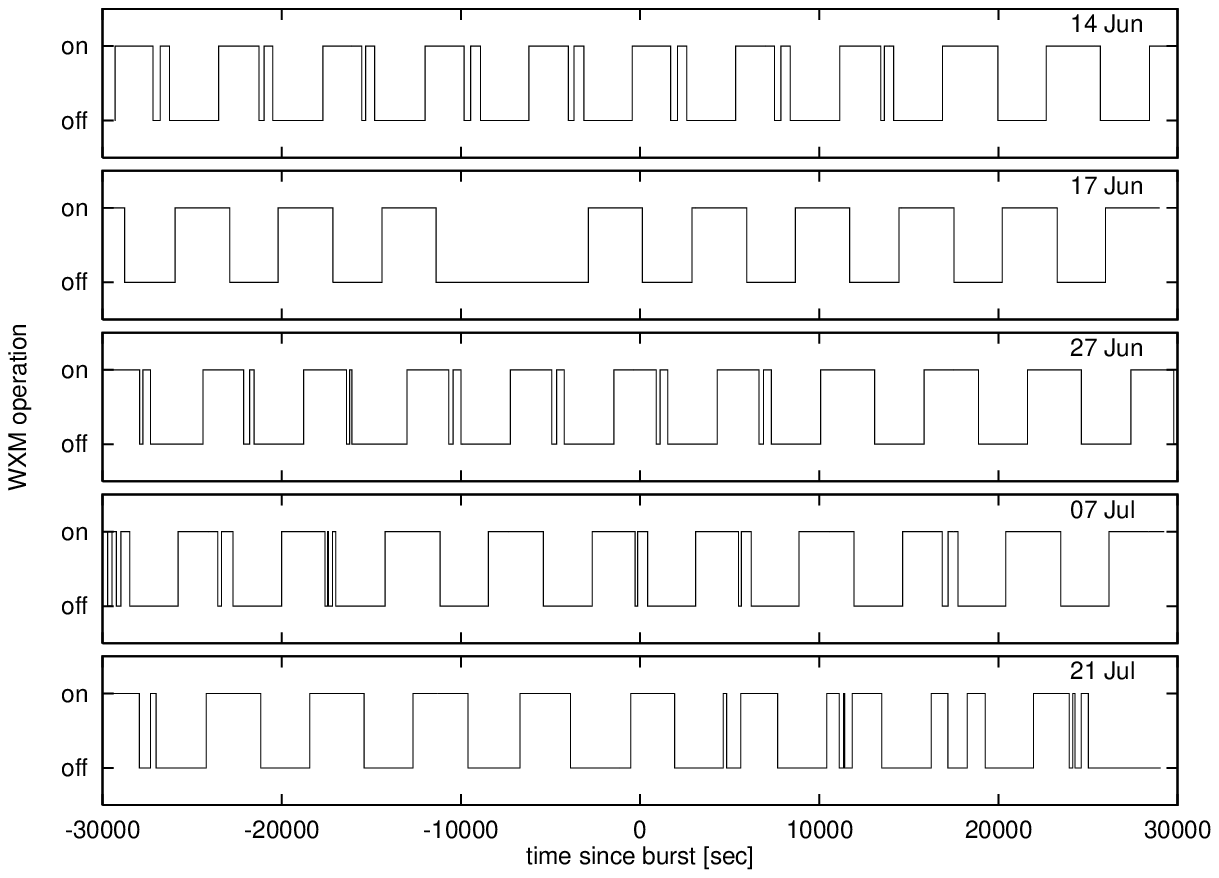}
 \end{center}
 \caption{The operation status of WXM around the burst
     on 14 June, 17 June, 27 June, 7 July, and 21 July 
     (from top to bottom).
     Each panel shows the status, on or off, of WXM
     from 30000 sec before the burst to 30000 sec after the bursts.
     The burst time are aligned at time=0.
     }
 \label{fig.orbits}
\end{figure*}

The WXM is sensitive to photons in 2$-$25 keV energy range.
Figure \ref{fig.orbits} shows the operation status (on or off) 
of WXM around the burst time.  
As shown in the figure, the observation efficiency of
WXM is approximately 50\%.
We use data of WXM for temporal and spectral analysis.
Table \ref{tab.obs} contains availability of the TAG data, which 
consist of time tagged photons with 32 energy bins.
The time resolution of the TAG data is precise enough ($<$ msec) to 
extract spectra of arbitrary time interval.
For the untriggered events (2nd and 3rd bursts), TAG data are not available.
So we use the PHA data, time-integrated 32-channel spectra
with 4.915 sec time resolution for the spectral analyses.
There are occasional dropouts of the TAG data.  Therefore, we use the 
TH data, time history in four energy bands
with 1.229 sec time resolution for the timing analyses in all bursts
in order to avoid the effects of the dropouts of the TAG data.

\subsubsection{Temporal Properties}
\begin{figure*}
 \begin{center}
    \FigureFile(59mm,){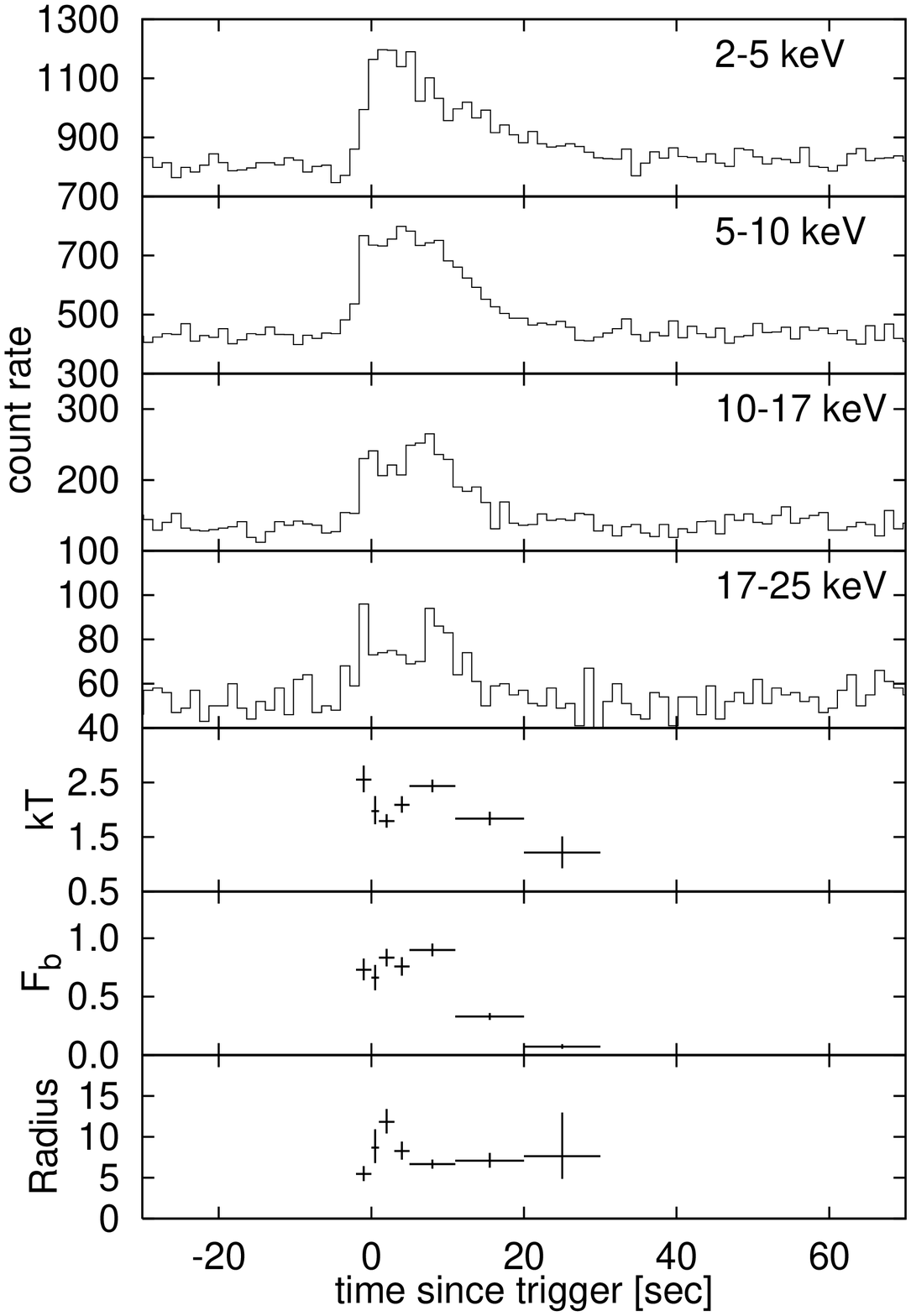}
    \FigureFile(59mm,){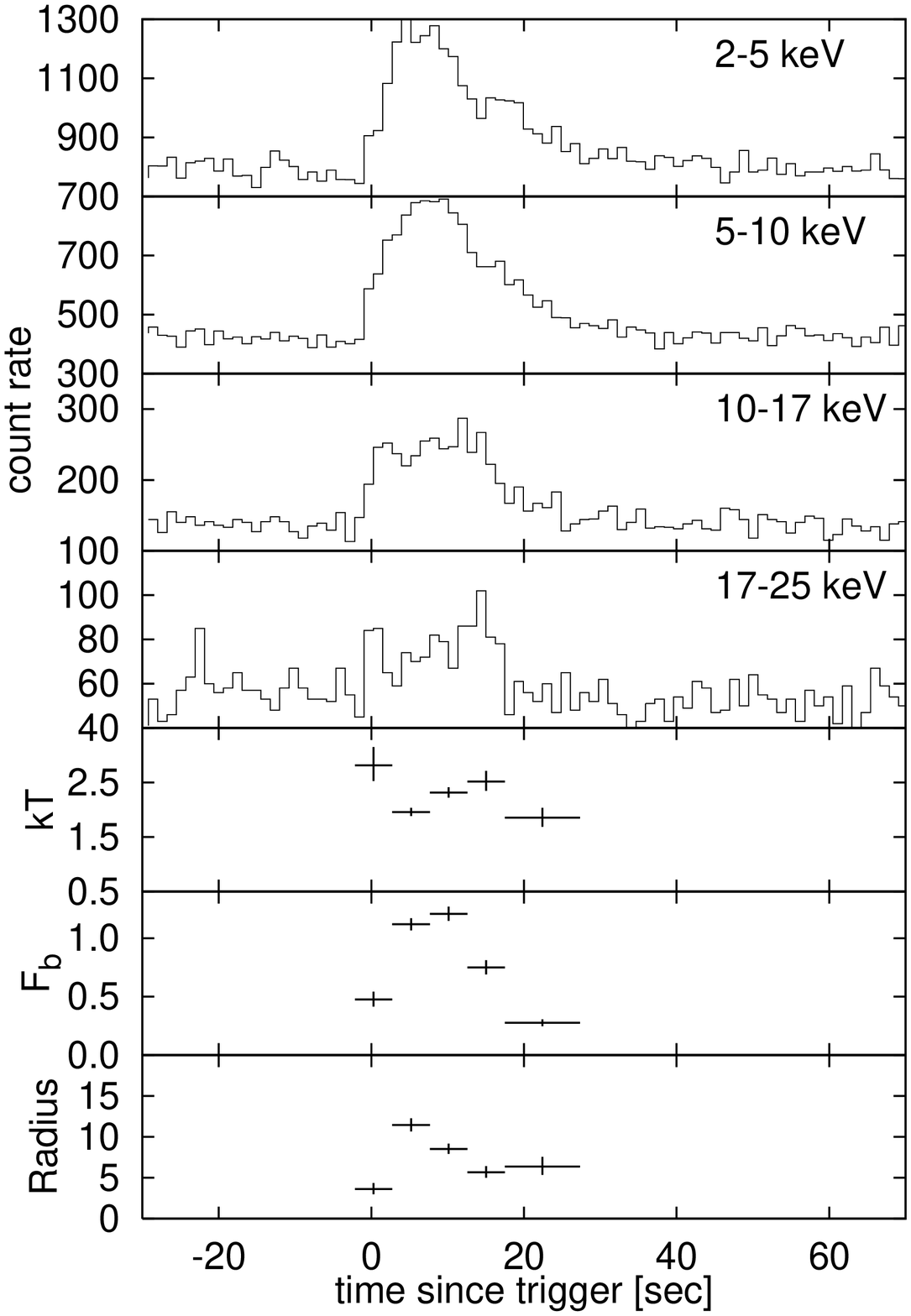}
    \FigureFile(59mm,){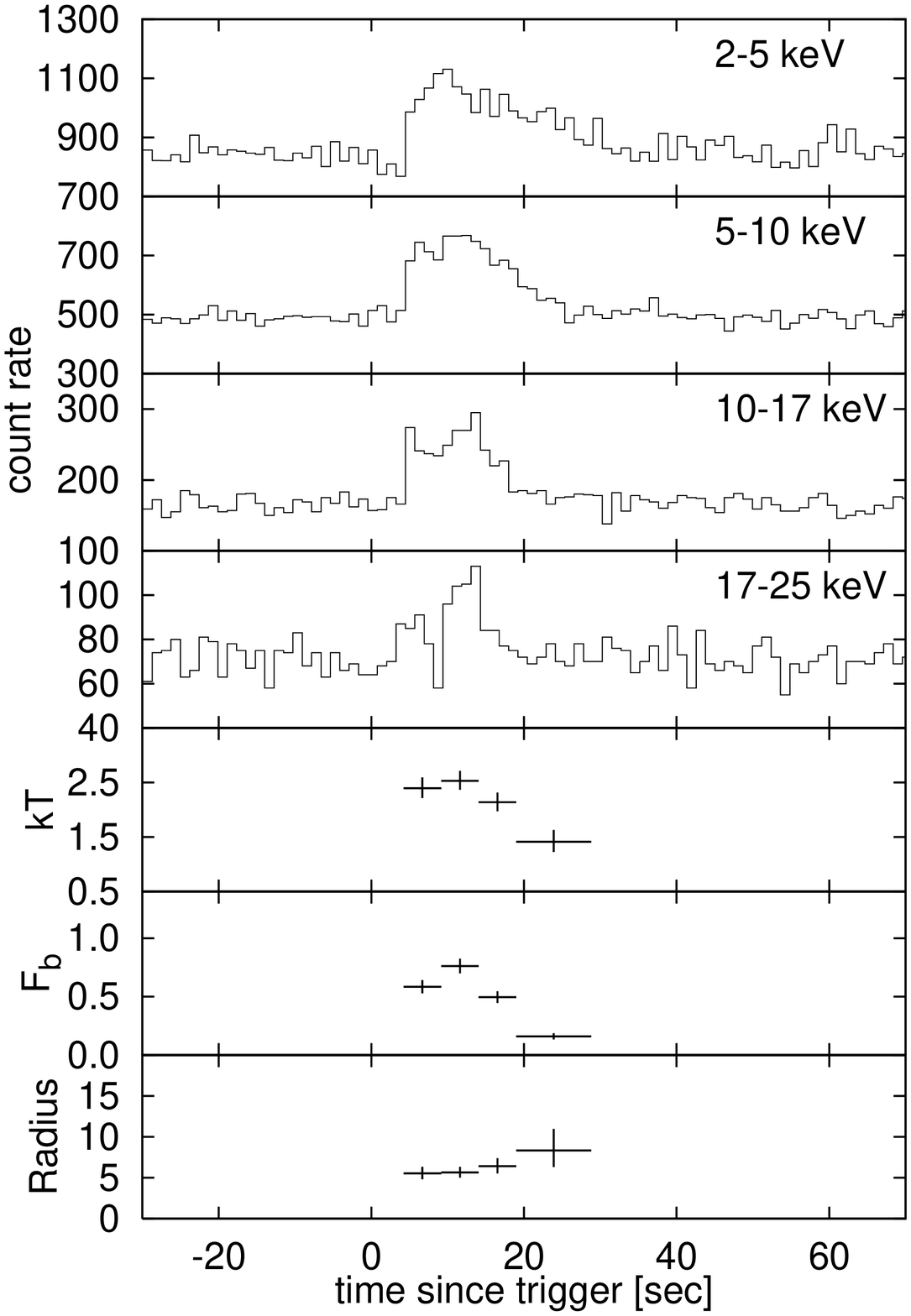}
    \FigureFile(59mm,){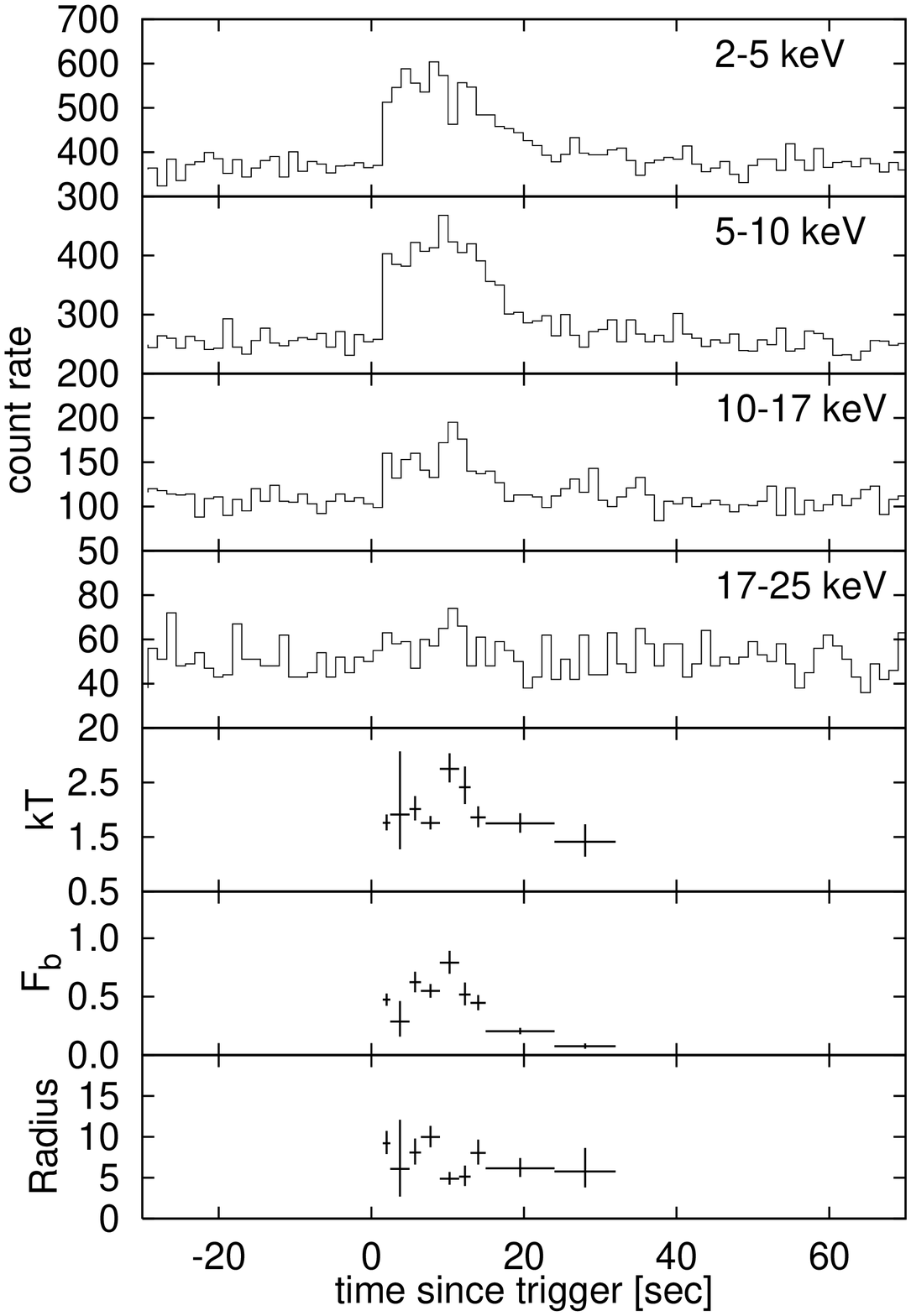}
    \FigureFile(59mm,){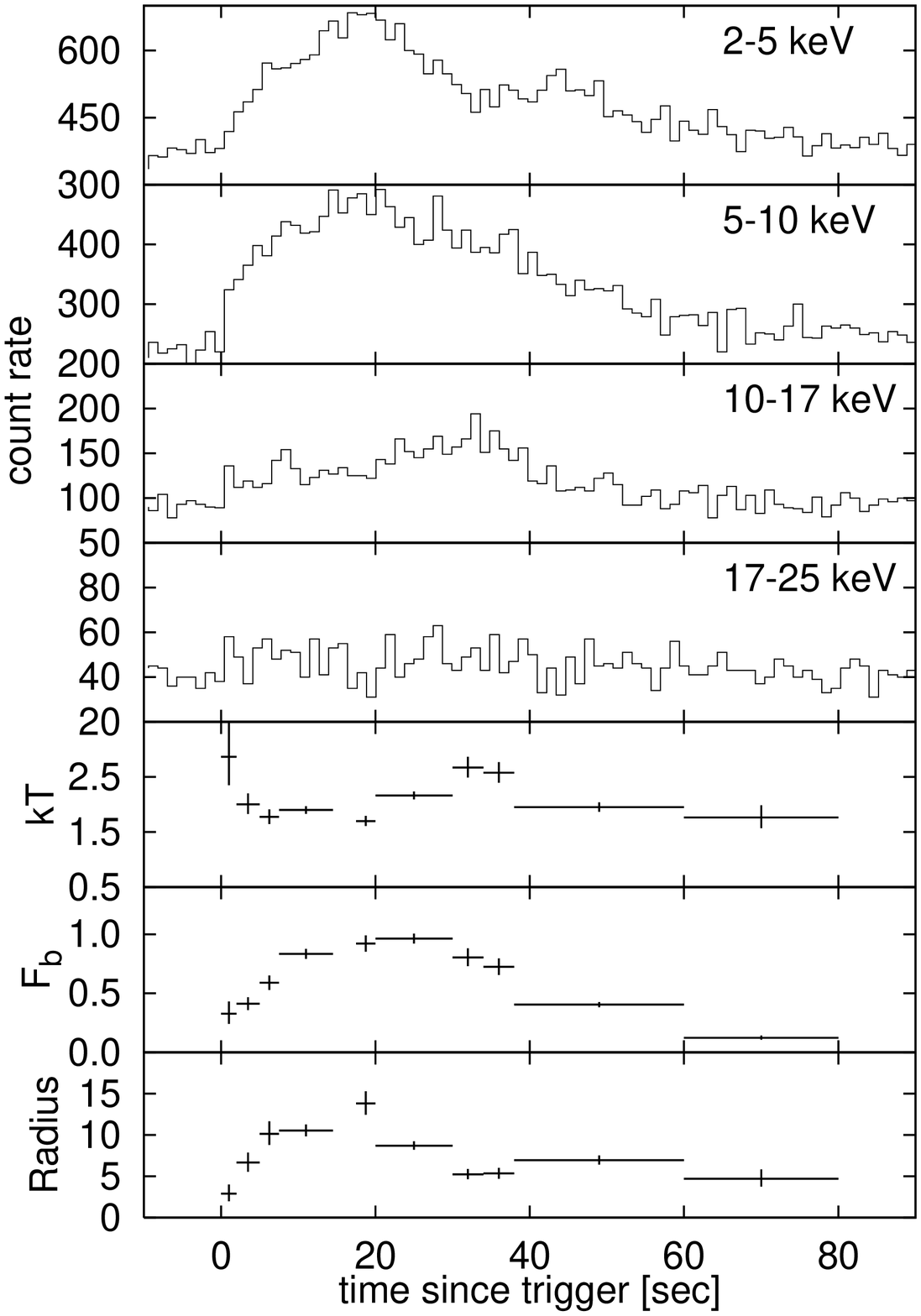}
 \end{center}
 \caption{The TH light curves of the X-ray burst HETE J1900.1-2455 in
     four energy bands observed by HETE-2 WXM.  
     The upper three plots are the bursts on 14 June, 17 June, and
     27 June (from left to right).
     The lower plots are the bursts on 7 July (left) and 21 July
     (right).  The count rates are in the unit of counts par
     1.229 sec bin.
     The spectral parameters of the fits to the blackbody model are 
     also shown in the lower three panels of each plots
     (see section \ref{section.spec}).
     The blackbody temperature $kT$ in the unit of keV, 
     the bolometric source flux $F_b$ in the unit of 10$^{-7}$ 
     ergs sec$^{-1}$ cm$^{-2}$, and the source radius in the 
     unit of km.  
     The distance to the source is assumed to be 4.31 kpc to
     estimate the radii of the blackbody radiation.
     The uncertainties are 90\% confidence limits. }
 \label{fig.lc}
\end{figure*}

Figure \ref{fig.lc} show the TH light curve of all five bursts
in 2$-$5, 5$-$10, 10$-$17, and 17$-$25 keV.  We can see the the double
peak structures in the 10$-$17 keV bands.  This features may be 
explained by an X-ray burst with moderate photospheric expansion
at the Eddington luminosity.

We calculated the duration $T_{90}$, which is a standard indicator
of the burst duration applied to gamma-ray bursts.
In Table \ref{tab.dur}, we summarized the $T_{90}$ 
duration of each burst in each energy band.
In the first four bursts, the $T_{90}$ durations are about 30 sec,
while the last burst has longer $T_{90}$, about 80 sec.

\begin{table*}
  \caption{Observed $T_{90}$
  durations of each 
  burst/energy band are summarized.  These durations are
  calculated using TH data in order to avoid the effects of
  the TAG data dropuouts.}
  \label{tab.dur}
  \begin{center}
  \begin{tabular}{lrrrr}
  \hline \hline 
  Burst No.& 2$-$5 keV & 5$-$10 keV & 10$-$17 keV & 17$-$25 keV \\
  \hline
  1 & 27.0 $\pm$  3.7 & 25.8 $\pm$ 10.6 
                         & 16.0 $\pm$ 37.1 & 16.0 $\pm$ 10.0 \\
  2 & 29.5 $\pm$  3.5 & 23.3 $\pm$  2.1 
                 & 23.3 $\pm$  2.7 & 18.4 $\pm$  8.8 \\
  3 & 23.3 $\pm$ 11.1 & 24.6 $\pm$  7.3 
                 & 17.2 $\pm$  7.2 & 29.5 $\pm$ 12.7 \\
  4 & 28.3 $\pm$  4.6 & 29.5 $\pm$  7.2 
                 & 28.3 $\pm$  8.9 & 47.9 $\pm$ 30.7 \\
  5 & 73.7 $\pm$  9.8 & 71.3 $\pm$  7.3 
                         & 50.4 $\pm$ 12.2 & 65.1 $\pm$ 17.3 \\
  \hline
  \end{tabular}
  \end{center}
\end{table*}

\subsubsection{Spectral Analysis}
\label{section.spec}

We performed the time resolved spectral analysis for the
bursts.  First, we divided the bursts into several time intervals
(see figure \ref{fig.lc}).  We used the data before and/or after
the burst as background data.  Although these background must contain
photons of the persistent emission from the source, they should be
negligibly small ($\sim 1/100$) compared with burst emission and
should not affect the results.
There were TAG data dropouts in the time interval between 14.5 sec and
17.5 sec of the 5th burst.  Therefore we did not analyze the data of
this time interval.  However according to the TH data,
the count rates and the hardness ratios of
this burst do not change much from 14.5 sec to 20.0 sec.
So the spectral parameters of the time interval of data lack may be
similar to that of the time interval between 17.5 sec and 20.0 sec.

We use the XSPEC version 11.3 software package for spectral fitting.
We find that the blackbody model is the best fit model
for the bursts.
We also tried the blackbody with photo-electric absorption model.
However we scarcely see the improvement in the $\chi^2$.
The N$_{\mathrm H}$ value is negligibly small.

We plotted the best fit spectral parameters with blackbody model
in figure \ref{fig.lc}: blackbody temperature $kT$
in the unit of keV, the bolometric source flux $F_b$
in the unit of 10$^{-7}$ ergs cm$^{-2}$ sec$^{-1}$, and the source
radius in the unit of km.  
In order to estimate source radius, we assumed the source distance
$d=4.3$ kpc, which is estimated from the
source flux and the Eddington luminosity (see section 
\ref{sec.distance}).

\begin{table*}
  \caption{The bolometric peak energy flux of each burst.
           The uncertainties are 90\% confidence limits.  }
  \label{tab.spec}
  \begin{center}
  \begin{tabular}{crcrr}
  \hline \hline 
  burst No. &
  \multicolumn{3}{c}{Time Region} & bolometric peak flux \\
  &
  \multicolumn{3}{c}{[s]} & [10$^{-7}$ ergs cm$^{-2}$ s$^{-1}$]
  \\ \hline
  1 &  5.0 &--& 11.0 & 0.90 $_{-0.05}^{+0.05}$ \\
  2 &  7.7 &--& 12.6 & 1.21 $_{-0.06}^{+0.06}$ \\
  3 &  9.2 &--& 14.1 & 0.76 $_{-0.06}^{+0.06}$ \\
  4 &  9.0 &--& 11.5 & 0.79 $_{-0.09}^{+0.10}$ \\
  5 & 20.0 &--& 30.0 & 0.96 $_{-0.04}^{+0.04}$ \\
  \hline
  \end{tabular}
  \end{center}
\end{table*}

\section{Discussion}

\subsection{The distance to the source}
 \label{sec.distance}
In section \ref{section.spec}, we showed that the spectra
of all time interval are consistent
with blackbody emission.
All the bursts have the time regions which show the drops of
blackbody temperature and the rises of source radius.
These features can be interpreted as photospheric
expansion at the Eddington luminosity.
The drop of blackbody temperature in the tail parts
show cooling.
All these results are consistent with Eddington-limited type-I
X-ray burst.

From the spectral fitting we obtained bolometric flux $F_b$ 
at the peak of each burst, and summarized in table \ref{tab.spec}.
The largest peak flux is that of the second burst, 
1.21 $\times 10^{-7}$ [ergs cm$^{-2}$ s$^{-1}$].
If it is equal to the Eddington luminosity of 1.4 solar mass neutron
star ($L = 2.7\times10^{38}$ [ergs s$^{-1}$]), 
the distance $d$ is
\begin{eqnarray}
   d &=&  \left(\frac{L}{4 \pi F} \right)^{1/2}\nonumber \\
        &=& 4.31 \, [\mathrm{kpc}] .
\end{eqnarray}

\subsection{The observation time and activities of the source}
WXM and FREGATE on HETE-2 detected only five X-ray bursts form 
HETE J1900.1-2455 in 2005 although the source was in the field of view
of HETE-2/WXM in every summer since 2001, as shown in the lower panel of
figure \ref{fig.asm}.
According to figure \ref{fig.asm}, we should have detected bursts between
2001 and 2004,  if the source had been active as much as in 2005. 
The absence of the burst before 2005 means
that the source was ``turned on'' some time after the summer of 2004.

In the insets of figure \ref{fig.asm}, the time of the bursts are
plotted with arrows.  
The figure clearly show the correspondence between persistent
flux and burst activity of the source, which is the common behavior
of bursts from the transient sources (e.g., 
Cen X-4: \cite{1980ApJ...240L.137M}; 
Aql X-1: \cite{1981ApJ...247L..27K}; 
X1608-522: \cite{1980ApJ...240L.143M,1989PASJ...41..617N}).
These observations agree with theoretical predictions
that bursts occur when the persistent flux is between 0.1\% and 10\%
of the Eddington luminosity
\citep{1978ApJ...220..291L,1978ApJ...225L.123J,1979Natur.280..375V}.
In the case of HETE J1900.1-2455, assuming the peak flux of the
bursts reached the Eddington luminosity, the persistent flux at the
time of bursts is about 1\% of the Eddington luminosity.  On the
other hand, the upper limit of the persistent flux in the quiescent
state is an order of magnitude smaller than in the active state,
which can be read from figure \ref{fig.asm}.  These facts are
consistent with the above theoretical expectation.

We can see the bursts occurred somewhat regularly in June and 
July of 2005 except the first burst.
Unfortunately, the WXM did not point toward the source after
early August.  Therefore we do not know the regularity of the
bursts after the last burst.
We should note here that the instruments on HETE-2
are turned off during the half of the orbits, and observation
efficiency is $\sim 50$\%.  There may be missing burst in the
time interval of regular activity.

We estimated $\alpha$ value \citep{1993SSRv...62..223L}, which
is ratio of the average persistent X-ray flux to the
average flux emitted in bursts, using the information of persistent
flux reported by \citet{2006ApJ...638..963K}.
The average persistent flux in 2$-$20 keV is $\sim 7 \times 10^{-10}$
erg cm$^{-2}$ sec$^{-1}$ during the observation.
To calculate average flux emitted in bursts $F_{\rm b}$, we divided
the burst fluence by the waiting time after the previous burst.
Since there may be some bursts we did not observed, $F_{\rm b}$
that we calculated should be lower limit.
The fluence of the 2nd, 3rd, 4th, and 5th bursts are $\sim$ 2.0, 1.1,
0.96, and 4.1 $\times 10^{-6}$ erg cm$^{-2}$ respectively.
The waiting time of these bursts are 2.97, 8.36, 8.61, and 12.4 
$\times 10^5$ sec.
Therefore upper limits of $\alpha$ for these bursts are $\sim$ 100, 500,
700, and 200.
The $\alpha$ for the 2nd burst is consistent with,
the standard $\alpha$ value for a pure helium atmosphere, which is $\sim$ 100.
It is unlikely that more than three missing bursts were present between
the waiting time, because the observation efficiency of WXM is
roughly 0.5.  Thus real $\alpha$ of the 3rd and 4th bursts might not be
lower than 150.  This $\alpha$ value may be slightly higher
than standard value.  
Here we note that
the estimated bolometric persistent flux may be smaller than the
real value,
because the persistent emissions of this type of sources have 
bright power-law component 
in addition to the blackbody emission from the neutron star surface. 
The harder and the brighter spectral shape makes the larger difference
between estimated and real bolometric persistent flux.  
However, the difference may not be larger than several tens percent
of the estimated flux. 
More precise consideration is possible in comparing the $\alpha$ value
of HETE J1900.1-2455 with other burst sources on the $\alpha$ and
$\gamma$ plane, where $\gamma$ is the bolometric persistent flux
normalized with the bolometric peak flux of the burst.
The studies of relation between $\alpha$ and $\gamma$ are summarized in
\citet{1988MNRAS.233..437V}, and there is a trend of lower $\alpha$ for
the source with lower $\gamma$.  On the other hand, our results,
$\log \alpha \sim 2$ and $\log \gamma \sim -2$, show high $\alpha$ 
despite low $\gamma$.  The other accreting millisecond pulsars
with type-I X-ray bursts, which are SAX J1808.4-3658
\citep{1998A&A...331L..25I,2001A&A...372..916I} and XTE J1814-338 
\citep{2003ApJ...596L..67S,2005ApJ...627..910K}, 
may show the same tendency.
Therefore, high $\alpha$ in low $\gamma$ might be a common property of
the accreting millisecond pulsar, while the confirmation with
larger samples is needed.

\subsection{The durations of the bursts}
The 5th burst had obviously longer duration than other four
bursts, and smaller $\alpha$ value at the same time.  
According to the model of hydrogen and helium flash, 
the burning of hydrogen together with helium makes the rise time
longer and $\alpha$ value smaller.  Since both short and long bursts
have the indications of photospheric expansion, both bursts
reach the Eddington limit.  Generally the Eddington limit corresponding
to the hydrogen-rich matter should be lower than the case of
the pure helium matter.  The observed peak flux are, however, 
almost constant or slightly higher in the longer burst.
These results are not fully understood in the framework of
hydrogen and helium flash model.

\section{Conclusion}
We discovered the new X-ray burst source HETE J1900.1-2455.
We analyzed five X-ray bursts from the source.
The temporal and spectral properties of the bursts are
consistent with those of Eddington-limited type-I bursts.
If the peak luminosity of the bursts is the Eddington luminosity of
a standard 1.4 solar mass neutron star with a helium atmosphere, the
distance to the source is $\sim$ 4 kpc.
The alpha values of the 3rd and 4th bursts, were estimated to be $>$150.
Such a large value can be explained by
the bright power-law component of the persistent emission.
The long duration and lower $\alpha$ value of 5th burst may
indicate the hydrogen-rich composition of burst fuel.
However the Eddington limit of the burst is comparable to the
other bursts.


\end{document}